\documentclass[12pt]{article}
\usepackage{psfig}
\usepackage{amsfonts}
 at 10truept
 at 10truept
 at 10truept
\def\be{\begin{equation}}
\def\ee{\end{equation}}
\def\ba{\begin{eqnarray}}
\def\ea{\end{eqnarray}}
\def\bq{\begin{quote}}
\def\eq{\end{quote}}

\def\ltap{\ \raise.3ex\hbox{$<$\kern-.75em\lower1ex\hbox{$\sim$}}\ }
\def\gtap{\ \raise.3ex\hbox{$>$\kern-.75em\lower1ex\hbox{$\sim$}}\ }
\def\gl{\ \raise.5ex\hbox{$>$}\kern-.8em\lower.5ex\hbox{$<$}\ }
\def\roughly#1{\raise.3ex\hbox{$#1$\kern-.75em\lower1ex\hbox{$\sim$}}}
\begin{document}

\thispagestyle{empty}
\begin{flushright}
LBNL-44717\\ UCB-PTH-99/55\\
hep-ph/9912453\\ \today
\end{flushright}
\vspace{1mm}
\begin{center}
\vspace*{1cm}
{\Large \bf SOLVING THE HIERARCHY PROBLEM WITH EXPONENTIALLY LARGE DIMENSIONS}\\
\vspace*{0.8cm}
{\large Nima Arkani-Hamed, Lawrence Hall, David Smith and Neal Weiner}\\
\vspace{1.5mm}
\vspace{0.5cm}
{\em
Department of Physics, University of California, Berkeley, CA 94530,
USA}\\
{\em Theory Group, Lawrence Berkeley National Laboratory,
Berkeley, CA 94530, USA}\\
\vspace{0.7cm}
ABSTRACT
\end{center}
In theories with (sets of) two large extra dimensions and supersymmetry
in the bulk, the presence of non-supersymmetric brane 
defects naturally induces a logarithmic potential for the volume of the 
transverse dimensions. 
Since the logarithm of the volume rather than the volume 
itself is the natural variable, parameters of O(10) in the potential can 
generate an exponentially large size for the extra dimensions.
This provides a true solution to the hierarchy problem, on the same 
footing as technicolor or dynamical supersymmetry breaking. 
The area moduli have a Compton wavelength of 
about a millimeter and mediate Yukawa interactions 
with gravitational strength. 
We present a simple explicit example of
this idea which generates two exponentially large dimensions. 
In this model, the
area modulus mass is in the millimeter range even for six dimensional 
Planck scales as high as 100 TeV.

\setcounter{page}{0} \setcounter{footnote}{1}
\newpage
\section{Introduction}
It has recently been realized that the fundamental scales of gravitational and
string physics can be far beneath $\sim 10^{18}$ GeV, in theories where 
the Standard Model fields live on a 3-brane in large-volume extra dimensions
\cite{ADD}. Lowering these fundamental scales close to the weak scale 
provides a novel 
approach to the hierarchy problem, and implies that the 
structure of quantum gravity may be experimentally accessible in the near 
future. 

While this prospect is very exciting, two important theoretical issues 
need to be addressed for this scenario to be as compelling as the 
more ``standard'' picture with high fundamental 
scale, where the hierarchy is stabilized by SUSY dynamically broken 
at scales far beneath the string scale. First: what generates the 
large volume of the extra dimensions? And second:
what about the successful picture of logarithmic gauge coupling unification
in the supersymmetric standard model? The success is so striking that we 
do not wish to think it is an accident.

One way of generating a large volume for the extra dimensions involves 
considering a highly
curved bulk. Indeed Randall and Sundrum have proposed a scenario where 
the bulk volume can be exponentially larger than the proper size of a single
extra dimension \cite{RS1}. Goldberger and Wise then showed how such a 
dimension could be stabilized \cite{Wise}. In the original proposal of
\cite{ADD}, however, the bulk was taken to be very nearly flat. Previous 
attempts at stabilizing large dimensions in this framework involved the 
introduction of large integer numbers in the theory, 
such as large topological charges 
\cite{Sundrum,ADMR1} or large numbers of branes \cite{ADMR1}. 
In this paper, we instead demonstrate how to stabilize  
{\it exponentially} large dimensions 
in the framework of \cite{ADD}. 

The set-up needed to accomplish this meshes nicely with recent discussions of 
how the success of logarithmic gauge coupling unification 
can be maintained with large dimensions and low string scale.
In \cite{Bachas,AntBac,AntBacDud,Ibanez,ADMR2} it was 
argued that logarithmic gauge 
coupling unification may be reproduced in theories with (sets of) two large 
dimensions. If various light fields propagate in effectively two transverse
dimensions, then the logarithmic Green's functions for these fields can give
rise to logarithmic variation of the parameters on our brane universe; 
in cases with sufficient supersymmetries, this logarithmic variation can 
exactly reproduce the logarithmic running of couplings seemingly far above the
(now very low) string scale. This phenomena is another example of the 
bulk reproducing the physics of the desert, this time with quantitative 
precision. Of course, for the ``infrared running'' picture to work after 
SUSY breaking, we must assume that SUSY is not broken in the bulk but only 
directly on branes. This is the analogue of softly breaking SUSY at low 
energies in the usual desert picture.

It is interesting that these same ingredients: sets of two transverse 
dimensions with SUSY in the bulk, only broken on branes, can also be 
used to address the issue of large radius stabilization. 
Indeed, in the SUSY limit, there is no bulk cosmological constant and there is
no potential for the radii; they can be set at any size. The crucial point 
is that once SUSY is broken on branes with a characteristic scale $\Lambda^4$,
locality guarantees that no bulk cosmological constant 
is induced, and therefore the
effective potential for the radius moduli does not develop any positive
power-law dependence on the volume of the transverse dimensions. For two 
transverse dimensions, logarithmic variation of light bulk fields can then 
give rise to a logarithmic potential for the size, $R$, 
of the extra dimensions:
\begin{equation}
V(R) \sim \Lambda^4 f(\mbox{log}(RM_*))
\end{equation}
where $M_*$ is
the fundamental scale of the theory.  
This can arise, for instance, from the infrared logarithmic variation of 
coupling constants on branes where SUSY is broken or from inter-brane forces
 \cite{AntBac,ADMR2,Gia}.
Since $\log (R)$ rather $R$ itself is the natural variable, 
if the potential has
parameters of $O(10)$, a minimum can result at log$({\overline R})
\sim 10$, thereby 
generating an exponentially large radius and providing a genuine solution 
to the hierarchy problem, on the same footing as technicolor or dynamical 
SUSY breaking. 

This idea is appealing and general; relying only
on sets of two transverse dimensions (for the logarithmic dependence) and 
supersymmetry in the bulk (to stably guarantee the absence of a bulk 
constant which would induce power-law corrections to the effective potential 
for the radii). It makes the existence of large extra dimensions seem 
plausible. However, the discussions in \cite{AntBac,ADMR2,Gia} have 
only pointed out this possibility on general grounds 
but have not presented concrete models realizing the idea. 
In this paper we remedy this situation by 
presenting an explicit example of a simple
theory with two extra dimensions, which stabilizes exponentially large 
dimensions. 
The interaction of branes with 
massless bulk scalar fields induces a logarithmic potential for the area
$A$ of the transverse dimensions of the form
\begin{equation}
V(A) = -f^4 + \frac{v^4}{\mbox{log}(AM_*^2)} + w^4 \mbox{log}(AM_*^2).
\end{equation}
This potential is minimized for an area 
\begin{equation}
{\overline A}M_*^2 = e^{v^2/w^2}
\end{equation}
and so only a ratio of $v/w \sim 6$ is needed to generate an 
area to generate the $\sim ($mm$)^2$ area 
needed to solve the hierarchy problem with $M_* \sim $ TeV. There is a 
single fine-tuning among the 
parameters $v,w$ and $f$, which are all of order $M_*$, 
to set the 4D cosmological constant to zero.

\section{The Radion Signal}
Since the potential for the radii of the extra dimensions vary only 
logarithmically, one might worry that the mass of the radius modulus about the
minimum of the potential will be too light. In fact, the mass turns out to 
be just in the millimeter range, and gives an observable deviation from 
Newton's law at sub-millimeter distances. 

Consider a 6 dimensional spacetime with metric of the form
\begin{equation}
ds^2=g_{\mu \nu}(x) dx^ \mu dx^ \nu + R^2 (x) \tilde{g}(y)_{mn} dy^m dy^n,
\label{eq:met1}
\end{equation}
where the geometry of $\tilde{g}$ is taken to be fixed at high energy scales; 
for example by brane configurations, as illustrated in the next
section. 
The low energy 4D effective field theory involves the 4D graviton
together with the radion field, $R(x)$, which feels the potential of
eq. (1). After a Weyl rescaling of the
metric to obtain canonical kinetic terms, the radion is found to have
a mass
\begin{equation}
m^2_R \sim 
\frac{R^2 V''(R)}{M_{Pl}^2} \sim \frac{\Lambda^4}{M_{Pl}^2} 
f''(\mbox{log}{R}) \sim \left(\frac{\mbox{TeV}^2}{M_{Pl}}\right)^2 \sim 
\mbox{mm}^{-2}.
\end{equation}
Hence, an interesting general consequence of such logarithmic potentials is that 
the mass of the radion is naturally
in the millimeter range for supersymmetry breaking and fundamental scales 
$\Lambda \sim M_* \sim $TeV. 
This order of magnitude result is important for mm range
gravity experiments, because the Weyl
rescaling introduces a gravitational strength coupling of the radion
to the Standard Model fields, so that radion exchange modifies the
Newtonian potential to
\begin{equation}
V(r) = -{G_N m_1 m_2 \over r} \; (1 + 2 e^{-m_R r}).
\label{eq:V}
\end{equation}
For a radion which determines the size of an $n$ dimensional bulk, the
coefficient of the exponential is $4n/(n+4)$, so that an observation of a 
coefficient corresponding to $n=2$ would be a dramatic signal of our
mechanism. 

It might be argued that, since $M_*$ is larger than 50--100 TeV for
$n=2$ from astrophysics and cosmology (\cite{CP,HS}), $m_R$ will
be sufficiently large that the range of the radion-mediated force will 
be considerably less than than a mm, making an experimental discovery 
extremely difficult. This conclusion is incorrect, for several reasons:
\begin{itemize}
\item The astrophysical and cosmological limits are derived from
graviton emission and hence constrain the gravitational scale, which
may be somewhat larger than the fundamental scale, $M_*$.
\item It is the scale of supersymmetry breaking on the branes,
$\Lambda$, which determines $m_R$, and this may be less than $M_*$,
reducing $m_R$ and making the range of the Yukawa potential larger.
\item The radion mass may be reduced from the order of magnitude
estimate $m_R \approx \Lambda^2/M_{Pl}$ by powers of log $R$, depending 
on the function $f$ which appears in the potential (1), 
as occurs in the theory described in the next section.
\item Finally, the cosmological and astrophysical limits on the
fundamental scale are \linebreak unimportant in the case that 
the bulk contains more than one 2D subspace, but as discussed in section 
4, the radions still have masses $\sim$ mm$^{-1}$.
\end{itemize}


\section{Explicit model}
In this section we present a specific effective theory that
stabilizes two large extra dimensions, without relying on input
parameters with particularly large ($>10$) ratios.  The framework
for our model is as follows.  Supersymmetry in the bulk guarantees
a vanishing bulk cosmological constant.  
Embedded in the 6D spacetime is a set of parallel
three-branes that can be regarded as non-supersymmetric defects.
Following closely the example of \cite{Sundrum}, the tensions of
these
three-branes themselves compactify the extra dimensions.  
We take the bulk bosonic degrees of freedom to be those of
the supergravity multiplet, namely, the graviton $g_{AB}$ and 
the anti-self-dual 2-form $A_{AB}$. 
The 2-form $A_{AB}$ does not couple to any of the three-branes and can 
be set to zero in our case. We can also have a set of 
massless bulk scalars $\phi_i$ contained in hypermultiplets. 
The relevant part of the Bosonic action is then
\begin{equation}
S = S_{Bulk} + S_{Brane}   
\end{equation}
where 
\begin{equation}
S_{Bulk} = 
\int d^4 x d^2 y \sqrt{-G}\left(-2 M^4 R + \sum_i (\partial \phi_i)^2 + \cdots
\right)
\end{equation}
is the bulk action and 
\begin{equation}
S_{Branes} = \int d^4 x \sum _i \sqrt{-g_i} \left(-f_i^4 + \sum_a 
{\cal L}_a(\psi_a,\phi|_a) + \cdots \right)
\end{equation}
is the action for the branes \cite{action}. Here the $f_i^4$ are the brane 
tensions and ${\cal L}_i$ are Lagrangians for fields $\psi_i$ that may 
live on the branes, which can also depend on the value of bulk fields 
$\phi$ evaluated on the brane $\phi|_a$. $G$ is the 6d metric,   
$g_i$ is the induced metric on the $i$'th brane, and we have set the 
bulk cosmological constant to zero. 

Note that while $S_{Bulk}$ must be accompanied by 
all the extra fermionic terms to have SUSY in the bulk, the brane actions
do not have to linearly realize SUSY at all, although they may realize SUSY 
non-linearly. In particular, there need not be any trace of superpartners 
on the brane where the Standard Model fields reside. The only reason
we need SUSY in the bulk is to protect against the generation of a bulk 
cosmological constant $\Lambda_{bulk}$, which would make a contribution 
$\sim \Lambda_{bulk} A$ to the potential for the area modulus and spoil our
picture with logarithmic potentials. 

Our model has three 3-branes, two
of which couple to scalars $\phi$ and $\phi'$.  
The dynamics on the brane impose boundary
conditions on the bulk 
scalar fields. In particular, imagine that the the brane
defects create brane-localized potentials for $\phi$, 
which want $\phi$ to take on the value $v_1^2$ on one 
brane and $v_2^2$ on the other. 
This will  
lead to a repulsive contribution to 
the potential for the area. 
The same two branes will be taken to have equal and opposite 
magnetic charges for the scalar $\phi'$, setting up a vortex-antivortex 
configuration for $\phi'$ which will lead to an attractive potential. 
The balance between 
these contributions provides a specific realization of how competing 
dependences on $\log R$ can lead to an exponentially 
large radius without very large or small input parameters.  

\begin{figure}
\centerline{
\psfig{file=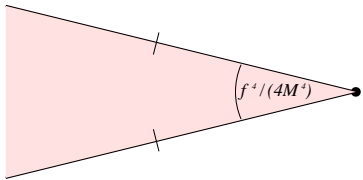,width=0.3\textwidth,angle=0}} 
\caption{The two transverse dimensions in the presence of
  a three-brane with tension $f^4$.  The shaded region is excluded, and
the two 
  borders of the excluded region are to be identified.}
\label{fig:onemass}
\end{figure}

\begin{figure}
\centerline{
\psfig{file=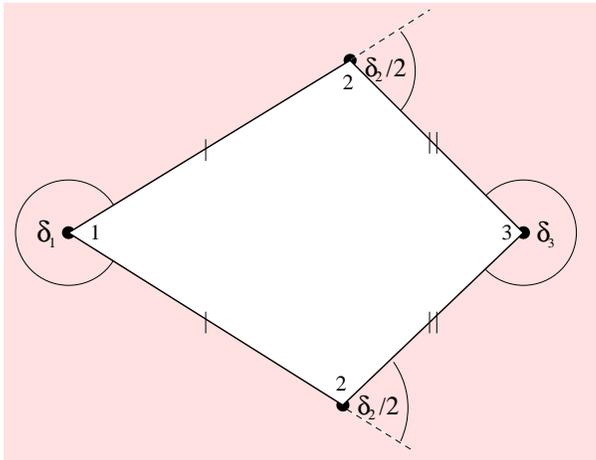,width=0.5\textwidth,angle=0}} 
\caption{A compact space can be obtained given three branes whose
  corresponding deficit angles $\delta_i$ 
 add up to $4\pi$.  Identifications to
  be made are indicated by hash marks.  Note that in contrast to 
  the brane in Figure \ref{fig:onemass}, 
branes 1 and 3 in this figure have tensions
larger than
  $4\pi M^4$.}
\label{fig:threemasses}
\end{figure}

We begin by reviewing how the brane tensions can compactify the two
extra dimensions \cite{action, Deser}.    
Suppose we
ignore for the time being the branes' couplings to bulk scalars, in
which case the relevant terms in the action in the low-energy limit are
\begin{equation}
S=-\int d^4 x \sum _i \sqrt{-g_i}  {f_i}^4-2M^4\int d^4 x d^2 y
  \sqrt{-G}  R.
\label{eq:action1}
\end{equation}
For the case in which
only a single brane is present, the static solution to Einstein's
equations 
is 
\begin{equation}
ds^2=\eta_{\mu \nu} dx^ \mu dx^ \nu + {\cal G}_{mn}(y) dy^m dy^n,
\label{eq:classicalg}
\end{equation}
where ${\cal G}_{mn}$ is the 2D Euclidean metric everywhere
but at the position of the three-brane, where it has a conical singularity
with deficit angle 
\begin{equation}
\delta ={f^4 \over 4M^4}.
\end{equation}
As expected, 
this is in exact correspondence with the metric around point masses in 
2+1 dimensional gravity \cite{Deser}. 
As shown in Figure \ref{fig:onemass}, the spatial dimensions
transverse to the brane are represented by the Cartesian plane
with a wedge of angle $\delta$ removed.
Adding a second brane removes a 
further portion of the Cartesian plane.  In fact, if ${f_1^4 \over 4M^4} +
{f_2^4 \over 4M^4}> 2 \pi$,
then the excluded region surrounds the allowed portion, as in Figure
\ref{fig:threemasses}.  
In this case
Einstein's equations have a static solution that features a compact space 
with spherical topology, {\em provided}
that a third brane of tension $f_3^4 = 16 \pi M^4 - f_1^4 - f_2^4$ is
placed at the intersecting lines of exclusion.
In general, a set of three-branes has a static solution with spherical
topology if
\begin{equation}
\sum_i {f_i^4 \over 4M^4}=4 \pi,
\end{equation}
that is, the deficit angles must add up to 4$\pi$.  

If a set of branes compactifies the space in this
manner, then the 4D effective theory is given by including in the
action of (\ref{eq:action1}) the massless excitations
about the classical metric.  Thus 
we replace $\eta_{\mu \nu} \rightarrow {\overline g}_{\mu \nu}(x)$ and
allow
${\cal G}_{mn}(y)$ to fluctuate about $\delta_{mn}$ in the bulk.  The
induced metric on a given brane will differ from ${\overline g}_{\mu
\nu}(x)$
by terms involving the fields associated with the brane separations,
which we temporarily ignore.  The curvature breaks up into two pieces
$R^{(4)}$ and $R^{(2)}$, the Ricci scalars built out of 
${\overline g}_{\mu \nu}(x)$ and ${\cal G}_{mn}(y)$, respectively.
Then, using the Gauss-Bonnet Theorem for spherical topology,
\begin{equation}
\int d^2 y \sqrt{{\cal G}}R^{(2)}=-8\pi,
\end{equation}
along with the fact that $R^{(4)}$ has no $y$ dependence,
we can integrate over the extra dimensions to obtain
\begin{equation}
S=-\int d^4 x \sqrt{-\overline{g}} \left( \sum_i f_i^4-16\pi M^4 +2(\int
d^2 y 
\sqrt{{\cal G}}) M^4 R^{(4)} \right).
\label{eq:action}
\end{equation}
In this action it is explicit that adjusting the
deficit angles to add up to $4\pi$ is equivalent to tuning the 4D 
cosmological constant to zero.

To develop our specific model we consider the case of three three-branes
on a space of spherical topology.  Then the ``shape'' of the extra
dimensions is fixed by the branes' deficit angles, or equivalently, by 
their tensions.  However, the size of the extra
dimensions,
\begin{equation}
A=\int d^2 y \sqrt{{\cal G}},
\end{equation}
is completely undetermined.  Moreover, the scalar 
associated with fluctuations of $A$, the radion, is massless and
mediates phenomenologically
unacceptable long-range forces.  
To stabilize the volume of the extra dimensions and give the radion a
mass, we
couple bulk scalar fields to two of the branes, which, for simplicity,
we assume have equal tensions $f$.
The scalar profiles will generate a potential $V_{\phi}(A)$ that
is minimized for a certain value $\overline A$ of the volume of the
compactified space.  
Adding the scalar action to (\ref{eq:action})
yields a total potential
\begin{equation}
V(A)=V_\phi(A)+\sum_i f_i^4  -16 \pi M^4.
\end{equation}
The effective cosmological constant,
\begin{equation}
\Lambda_{eff}=V(\overline{A}),
\label{eq:lambdaeff}
\end{equation}
can then be made to vanish by a single fine tuning of fundamental
parameters.  The
back-reaction on the spatial geometry that is induced by the scalars
is discussed below.

We work with two massless bulk scalars, $\phi$ and $\phi'$, which
induce repulsive and attractive forces, respectively. 
In treating the scalar fields, we will for 
simplicity ignore their back-reaction on the 
metric and assume that they propagate in the flat background with conical 
singularities set up by the branes. It is easy to see that the effect of 
back-reaction can be made parametrically small if the scalar energy scales 
are somewhat smaller than $M_*$, and none of our conclusions are affected. 

\begin{figure}
   \centerline{
\psfig{file=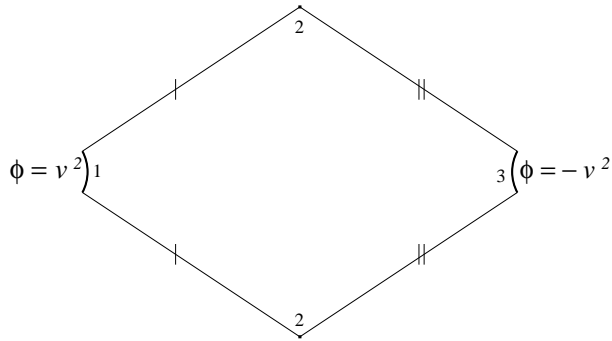,width=0.5\textwidth,angle=0}} 
\caption{The boundary conditions on $\phi$.  Identifications to be
  made are indicated by hash marks.}
\label{fig:fullspace}
\end{figure}

Suppose that on branes 1 and 3 of Figure \ref{fig:fullspace}, 
$\phi$ is forced to take on unequal values $v^2_1$ and $v^2_3$,
respectively. This can for instance be enforced if the non-SUSY brane defects
generate a potential for $\phi$ on the branes, analogous to what was 
considered in \cite{Wise}.  
Because $\phi$ is massless in the bulk, 
we are free to perform a constant field
redefinition and take $v^2_1 = -v^2_3 \equiv v^2$. We account
for the brane thicknesses by enforcing these values for $\phi$ to hold
along arcs of
finite radius $r_* \sim 1/M_{*}$, and not just at individual points.  The
field
configuration in the bulk is then given by solving Laplace's
equation with these boundary conditions.

\begin{figure}
\centerline{
\psfig{file=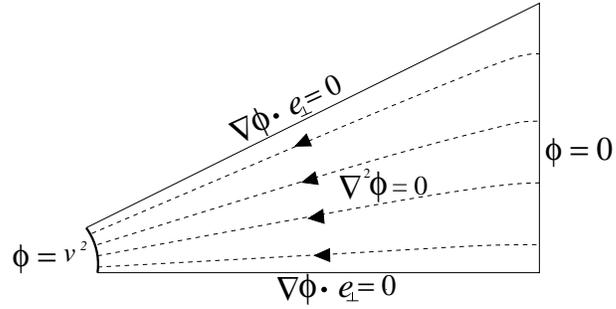,width=0.5\textwidth,angle=0}} 
\caption{A boundary-value problem that determines $\phi$.  Here ${\mathbf
  e}_\bot$ refers to the unit vector normal to the relevant boundary,
and lines of ${\mathbf \nabla \phi}$ are shown dashed.
The solution for the full space of Figure \ref{fig:fullspace} is given by
first 
evenly reflecting across the bottom horizontal line, and then performing 
an odd reflection
(i.e., $\phi \rightarrow -\phi$) across the vertical line where
$\phi=0$.}
\label{fig:quarterspace}
\end{figure}

Keeping in mind the identifications to be made between the various
edges of the space in Figure \ref{fig:fullspace}, the symmetry of the
problem tells us
that the field configuration is found by solving the problem
depicted in Figure \ref{fig:quarterspace}, and then reflecting that
solution appropriately.
For simplicity we consider instead a slightly different problem which,
unlike that shown in Figure \ref{fig:quarterspace},
is trivially solved.  
\begin{figure}
\centerline{
\psfig{file=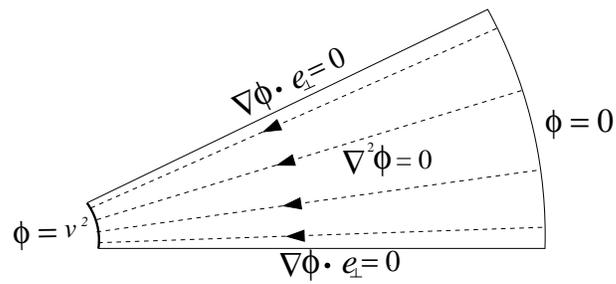,width=0.5\textwidth,angle=0}} 
\caption{The simplified boundary-value problem for $\phi$.}
\label{fig:pie}
\end{figure}
As indicated in Figure \ref{fig:pie}, 
we take the boundary at which $\phi=0$ holds to be an
arc of radius $R$, rather than a straight line, so that the solution
in this region is immediately found to be
\begin{equation}
\phi=v^2 {\log{(R/ r)}\over \log{(R/ r_*)}},
\end{equation}
where $r$ measures the distance from the (missing) left vertex of the
pie slice.  The total energy of this configuration is 
\begin{equation}
4 \int d\theta \int_{r_*}^R dr r  {({\mathbf \nabla}
  \phi)^2 \over 2}  = \theta_0 {v^4 \over  \log{(R/ r_*)}},
\label{eq:repulsive}
\end{equation}
where $\theta_0=2\pi-{f^4
  \over 4 M^4}$.  
Thus, $\phi$ sets up a
$1/\log{R}$ repulsive potential.  It is not difficult to prove using
simple variational arguments that
the same conclusion is reached when one solves the ``real'' problem
involving the triangle rather than the pie slice.

Now suppose that the same two branes that couple to $\phi$ carry
topological charge under 
a derivatively coupled field $\phi'$.  That is, under any closed loop
containing a brane we have
\begin{equation}
\int {\mathbf d} {\mathbf l} \cdot {\mathbf \nabla} \phi'= n \theta_0 w^2,
\end{equation}
where $w$ is a fixed parameter with unit mass dimension and $n$ is an integer.
Non-zero charge $n \neq 0$ is only possible if we make the 
identification
\begin{equation}
\phi \sim \phi +\theta_0 w^2.
\end{equation}
In order to be able to solve Laplace's equation on a compact space, the branes
must carry equal and opposite charges, which we take to correspond to 
$n= \pm 1$. 
The configuration for $\phi'$ is then found by
solving Laplace's equation with
${\mathbf \nabla} \phi'=\pm {w^2 \over r_*} \mathbf{e_\|}$ on the branes
(the gradient runs clockwise on one brane and counterclockwise on
the other).
This sets up the the vortex-antivortex field configuration for 
$\phi'$ shown in Figure 6.
\begin{figure}
\centerline{
\psfig{file=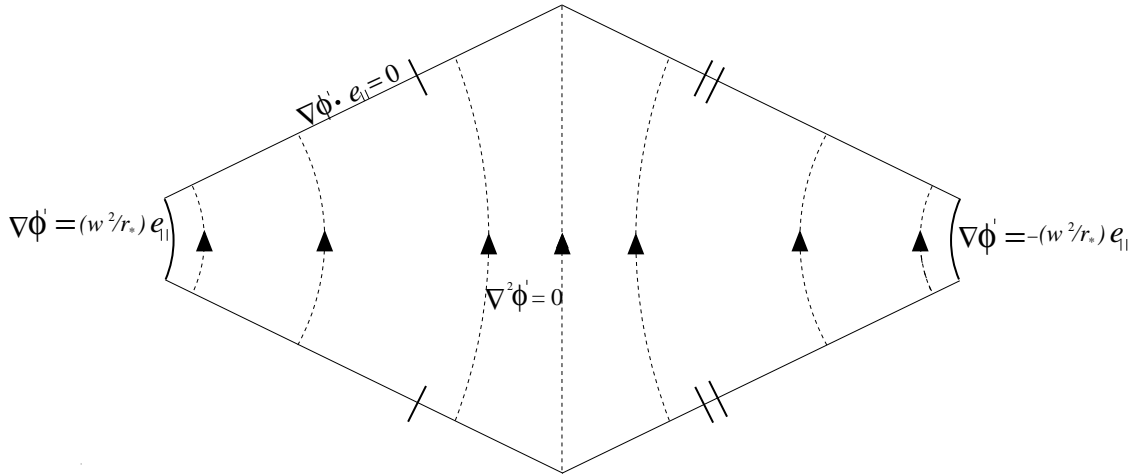,width=0.93\textwidth,angle=0}} 
\caption{The configuration of $\phi'$. Each brane carries a
  topological charge, which generates an attractive potential.}
\label{fig:gradreal}
\end{figure}
For simplicity, in order to calculate the energy in this configuration we 
once again work on a pie slice (Figure \ref{fig:gradpie})  
\begin{figure}
\centerline{
\psfig{file=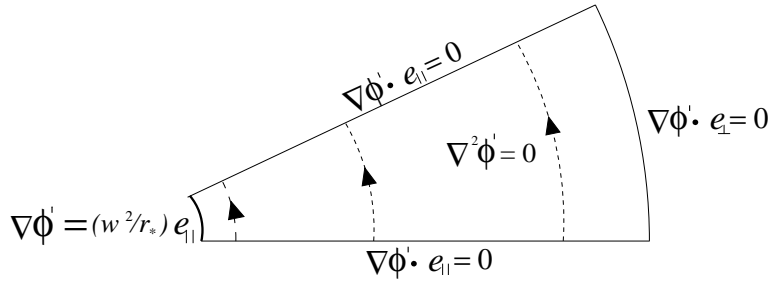,width=0.63\textwidth,angle=0}} 
\caption{The simplified boundary-value problem for $\phi'$ .  Here
  $\mathbf{e_\bot}$ and $\mathbf{e_\|}$ are the unit vectors normal and
  parallel, respectively, to the relevant boundary. }
\label{fig:gradpie}
\end{figure}
rather than a triangle, 
and it is easily proved
that this modification does not affect the essential scaling of the
energy with the area. With this simplification the solution is
\begin{equation}
\phi'=w^2 \theta +C,
\end{equation}
where $\theta$ is the angular coordinate and $C$ is an
undetermined, irrelevant constant.  The energy of the
configuration is then found to be
\begin{equation}
4 \int d\theta \int_{r_*}^R dr r  {({\mathbf \nabla}
  \phi')^2 \over 2} = \theta_0 w^4 \mbox{log}(R/r_*).
\label{eq:attract}
\end{equation}
so we have found an attractive potential that will balance the
repulsive contribution of (\ref{eq:repulsive}).
>From (\ref{eq:repulsive}) and (\ref{eq:attract}), we see that the 
full potential is 
\begin{equation}
V(R) = \theta_0 {v^4 \over \log(R/r_*)} + \theta_0 w^4 \log(R/r_*)
+\sum_i f_i^4  -16 \pi M^4,
\label{eq:totalpotential}
\end{equation}
which is minimized when
\begin{equation}
R=\bar{R} = r_* e^{v^2/w^2}.
\end{equation}
Even a mild ratio $v/w \sim 6$ yields an exponentially large 
radius $\bar{R} \sim 10^{16} r_*$.  
The effective cosmological constant,
\begin{equation}
\Lambda_{eff}=V(\bar{R}) = \sum_i f_i^4 -16 \pi M^4 +2 \theta_0 v^2 w^2,
\end{equation}
can be made to vanish by a single tuning of $v$, $w$, and the
brane tensions.

Note that we can now see explicitly that the presence of the non-supersymmetric
brane defects can not generate a bulk cosmological constant. 
The presence of the branes leads to logarithmic variation for the
bulk fields, which does indeed break SUSY and generate a potential for the 
area modulus. However, since any {\it constant} field configuration preserves
SUSY, the SUSY breaking in the bulk must be proportional to the {\it gradient}
of the bulk scalar fields, which drops as $1/r$ with distance $r$ away from 
the branes. Therefore, it is impossible to induce a cosmological constant,
since this would amount to an constant amount of SUSY breaking throughout the
bulk. In fact, a very simple power-counting argument shows that all corrections
to the energy are logarithmic functions of the area. 

Given a specific form for the logarithmic potential
(\ref{eq:totalpotential}), we can work out the 
mass of the area modulus, which is 
\begin{equation}
m^2_{R} \sim \frac{\bar{R}^2 V''(\bar{R})}{M_{Pl}^2} \sim \frac{v^4}{M_{Pl}^2 
\mbox{log}^3(\bar{R}/r_*)}.
\end{equation}
Interestingly, $m_R$ is suppressed
by (log$(\bar{R}/r_*))^{3/2}$ 
compared to the naive estimate $M_*^2/M_{Pl}$. Hence
even for $v \sim M_*$ as large as 100 TeV, the range of the radion-mediated
Yukawa potential is 0.1 mm -- accessible to planned experiments.

\section{Four and Six Extra Dimensions}
Since the logarithmic form of the propagator
occurs only in two dimensions, one may worry that the ideas in this paper 
are only applicable to the case of two large dimensions. This is the 
case most severely constrained by astrophysical and cosmological constraints 
\cite{ADD,CP,HS}, which demand the 6D Planck scale $M_* >$ 50 TeV, seemingly
too large to truly solve the hierarchy problem. One possibility 
is that the true Planck
scale of the ten dimensional theory could be $\sim$ O(TeV),
and the 6D Planck scale of $\sim 50$ TeV could arise if  
the remaining four dimensions are a reasonable factor 
$O(10)$ bigger than a (TeV$)^{-1}$. But we don't have to resort to this 
option. As pointed out in \cite{AntBac,ADMR2}, the presence of two-dimensional
subspaces where massless fields can live is sufficient to generate logarithms.
Take the case of four extra dimensions. Imagine one set of 
parallel 5-branes filling out the 12345 directions, and another set 
filling out the 12367 directions. They will intersect on 3-dimensional spaces
where 3-branes can live. These 3-branes can act as sources for fields
living on each of the 5-branes, which effectively propagate in two sets of 
orthogonal 2D subspaces. 
Once again, bulk SUSY can guarantee a vanishing  
``cosmological constant'' for each of the 2D subspaces. The  SUSY breaking
at the intersections can set up logarithmically 
varying field configurations on the 5-branes
that leads to a potential of the form  
$V(\mbox{log}A_1,\mbox{log}A_2)$ for the areas $A_1,A_2$ of the 2D subspaces. 
Minimizing the potential, each radius can be exponentially large, 
and the ratio of the radii will also be exponential, but the
value of $M_{Pl}$ will require the largest
radius to be very much smaller than a mm. It would be interesting to build 
an explicit model along these lines. 

Even without an explicit model, however, we can see that 
the scale of the radion masses is unchanged. The logarithmic potential  
still gives $m_{Ri} \approx \Lambda^2/M_{Pl}
\approx \mbox{mm}^{-1}$, for $\Lambda \approx 1$ TeV. After Weyl rescaling,
each radion couples with gravitational strength to the Standard Model and 
should show up in the sub-millimeter measurements of gravity. 

\section{Other ideas}
There is an alternative way 
in which theories with two transverse dimensions can generate 
effectively exponentially large radii. 
The logarithmic variation of bulk fields can force the theory
into a strong-coupling region exponentially far away from some branes,
and interesting physics can happen there. 
This is the bulk analog of 
the dimensional transmutation of non-Abelian gauge theories, which generate
scales exponentially far beneath the fundamental scale and trigger interesting
physics, such as e.g. dynamical supersymmetry breaking \cite{ADMR2}. It is 
tempting to speculate that such strong-coupling behavior might effectively 
compactify the transverse two dimensions. Recently, Cohen and Kaplan have
found an explicit example realizing this idea \cite{CK}. They consider a massless scalar
field with non-trivial winding in two transverse dimensions: a global 
cosmic string. Since the total energy of the string diverges logarithmically
with distance away from the core of the vortex, we expect gravity to become
strongly coupled at exponentially large distances. Indeed, Cohen and Kaplan
find that the metric develops a singularity at a finite proper distance from 
the vortex core, but argue that the singularity is mild enough to be rendered 
harmless. What they are left with is a non-compact transverse space, with 
gravity trapped to an exponentially large area
\begin{equation}
\bar{A}M_*^2 = e^{M_{*}^4/f_{\pi}^4}
\end{equation}
where $f_\pi$ is the decay constant of the string. A ratio of $M_{*}/f_{\pi} 
\sim 2.5$ is all that is needed to solve the hierarchy problem in this case. 
This model is a natural implementation of the ideas of \cite{ADD}, 
to solve the 
hierarchy problem with large dimensions, together with the idea of trapping
gravity in non-compact extra dimensions as in \cite{RS2}. Unlike \cite{
RS1}, however, the bulk geometry is not highly curved everywhere, 
but only near the 
singularity. Thus, gravity has essentially been trapped to a flat ``box'' of
area $A$ in the transverse dimensions, and the phenomenology of this scenario 
is essentially the same as that of \cite{ADD}. An attractive aspect of this 
scenario is that, unlike both our proposal in this paper 
and those of \cite{RS1,Wise}, 
no modulus needs to be stabilized in order to solve the hierarchy problem.
This also points to a phenomenological 
difference between our proposal and that of \cite{CK}. While both schemes
generate an exponentially large area for two transverse dimensions, there 
is no light radion mode in \cite{CK} whereas we have a light radion with 
$\sim$mm$^{-1}$ mass.

\section{Conclusions}
In this paper, we have shown how to stabilize 
exponentially large compact dimensions, 
providing a true solution to the hierarchy problem 
along the lines of \cite{ADD} which is on the same footing as technicolor and 
dynamical SUSY breaking. Of course, there are many mysteries other 
than the hierarchy problem, and the conventional picture of beyond the Standard
Model physics given by SUSY and the great desert had a number of successes. 
So why do we bother pursuing alternatives? Are 
we to think that the old successes are just an accident? 

A remarkable feature of theories with large extra dimensions is that 
the phenomena that used to be understood inside the energy desert can also 
be interpreted as arising from the space in the extra dimensions. Certainly
all the qualitative successes of the old desert, such as 
explaining neutrino masses and proton stability, can be exactly 
reproduced with the help of the bulk \cite{ADD,Neu,AS,AD}, in such a way that 
e.g. the success of the see-saw mechanism in explaining the scale of neutrino 
masses is not an accident. 
As we have mentioned, there 
is even hope that the one quantitative triumph of the 
supersymmetric desert, logarithmic gauge coupling unification, 
can be exactly reproduced so that the old success is again not accidental.
We find it encouraging that it is precisely the same sorts of models-with two 
dimensional subspaces, SUSY in the bulk broken only on branes- which allows
us to generate exponentially large dimensions. Hopefully, in the next
decade experiment will tell us whether any of these ideas are relevant to
describing the real world.

\vskip 0.25in
\appendix
\noindent {\bf \Large Acknowledgements}
\vskip 0.15in
\noindent This work was supported in part by the Director, Office of Science,
Office of High Energy and Nuclear Physics, Division of High Energy
Physics of the U.S. Department of Energy under Contract
DE-AC03-76SF00098 and in part by the National Science Foundation under
grant PHY-95-14797. 

\end{document}